\documentclass[11pt,journal,final,a4paper,twoside,compsoc]{IEEEtran}
\usepackage[T1]{fontenc}
\usepackage[utf8]{inputenc}
\usepackage{amssymb}
\usepackage{cite}
\usepackage{flushend}
\usepackage{fontawesome}
\usepackage{graphicx}
\graphicspath{{../img/}}
\usepackage[hidelinks]{hyperref}
\usepackage{textcomp}
\usepackage{xcolor}

\definecolor{insightcol}{rgb}{0.6,0,0}
\definecolor{conjecturecol}{rgb}{0.2,0.0,0.0}
\definecolor{quotecol}{rgb}{0,0.6,0.2}
\definecolor{nonquotecol}{rgb}{0,0.4,0.1}
\definecolor{respondentcol}{rgb}{0,0,0.7}
\definecolor{ivcol}{rgb}{0,0,0.7}

\newcounter{insight}
\renewcommand{\theinsight}{\arabic{insight}}
\newcounter{conjecture}
\renewcommand{\theconjecture}{\arabic{conjecture}}
\newcounter{evidence}[insight]

\newcommand{\IV}[1]{\textcolor{ivcol}{{``#1''}}}
\newcommand{\R}[1]{\textcolor{respondentcol}{\textsuperscript{R#1}}}
\newcommand{\rQuote}[2]{\textcolor{quotecol}{``#2''}\R{#1}\index{R#1}}

\newcommand{\Insight}[2]{\refstepcounter{insight}\label{#1}
  \vspace{1ex}\par\noindent
  \bgroup \color{insightcol}\bfseries Insight \theinsight: #2 \egroup}
\newcommand{\Conjecture}[2]{\refstepcounter{conjecture}\label{#1}
  \vspace{1ex}\par\noindent
  \bgroup \color{conjecturecol}\bfseries Conjecture \theconjecture:
  \color{insightcol} #2 \egroup}

\newcommand{\citep}[1]{\cite{#1}}

\begin{document}

\title{Does ICSE Accept the Right Contributions?}

\author{\IEEEauthorblockN{Lutz Prechelt}
  \IEEEauthorblockA{\textit{Freie Universit\"at Berlin} \\
  Berlin, Germany \\
  prechelt@inf.fu-berlin.de}
}

\maketitle

\begin{abstract}
\emph{Background:} 
There is a constant discussion regarding whether the ICSE Technical Research
track is accepting too many contributions of some type and
too few of some other type.\\
\emph{Questions:}
Are ICSE and the contributions it is seeing well aligned with
what is important for bringing software engineering forward?\\
\emph{Method:}
26 expert interviews with senior members of the ICSE community,
evaluated qualitatively and reported with many quotations.\\
\emph{Results:}
About three quarters of the respondents are not generally happy with
ICSE's alignment.
Two specific complaints that recur frequently concern
a) many low-relevance contributions making it into the program and
b) several types of high-relevance contributions hardly seen in
the ICSE program.
\end{abstract}

\section{Background}

ICSE, the International Conference on Software Engineering,
established in 1978,
is widely considered the top conference for software engineering research.
Due to its high prestige, researchers from all areas of
software engineering research are constantly pressing to get their work
into the conference and the ICSE steering committee
(in particular each year's program chairs)
is constantly required to justify the way in which they balance 
the program \cite{BriaHoe14}.

Over time, this has led to many changes (and subsequent reversal of some
of them) in the way ICSE is structured and how it makes its decisions.
For instance, there are currently four additional tracks besides the
main (``Technical Research'') track, SEIP, SEIS, SEET, and NIER,
plus presentations of journal-first papers in an attempt to get a broader
selection of relevant research into the conference.
However, the additional tracks have far lower prestige than the
main track and so the pressure on the main track continues to find an answer
to the question ``Are we accepting the right works?''.

Many ICSE participants have a position on this question,
but little of that is ever reflected in written form in public places.
As a by-product of an interview study I ran at ICSE 2018,
I collected some data in this regard.
Zhe present (rather informal) report is meant to make a summary of it
publicly available.
The informality means I will only sketch the method used
and present the observations (with many quotes),
but make no attempt to discuss related work, limitations, or conclusions.

\section{Method}

\begin{figure}
  \includegraphics[width=\columnwidth]{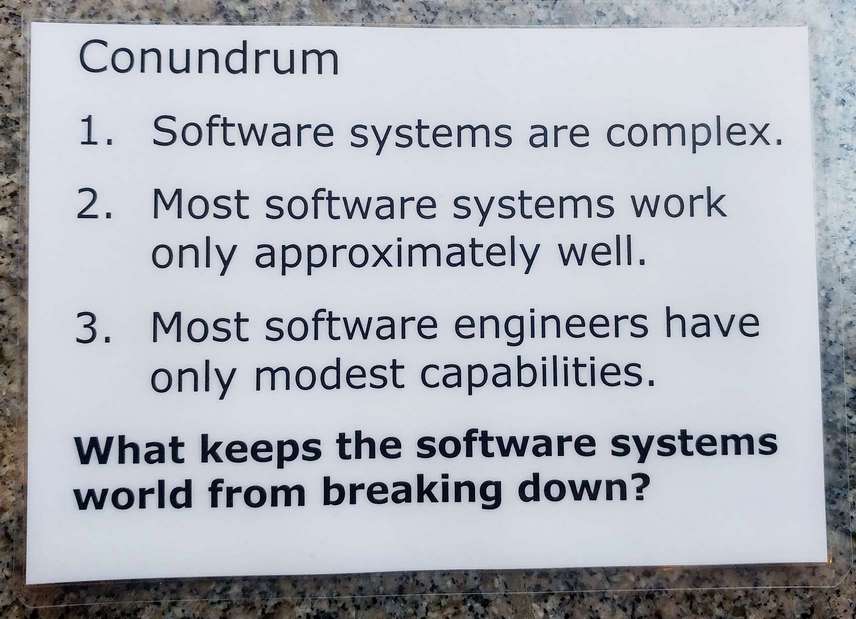}
  \caption{Stimulus card used in the interviews}\label{stimulus.jpg}
\end{figure}

The interview study concerned the provocative question 
``What keeps the software systems world from breaking down?''
and was mostly driven by the 4-item stimulus card shown in
Figure~\ref{stimulus.jpg}.

It contained three statements S1, S2, S3 meant to establish a context for 
the main interview question QQ.
\begin{itemize}
\item \textbf{S1}: Software systems are complex.
\item \textbf{S2}: Most software systems work only approximately well.
\item \textbf{S3}: Most software engineers have only modest capabilities.
\item \textbf{QQ}: What keeps the software systems world from breaking down?
\end{itemize}

The method details of these interviews and their analysis
and the outcomes of that analysis are described elsewhere
\cite{Prechelt19-knowgaps}.

On this basis, I ran 54 semi-structured interviews with
respondents from 18 different countries.
16 respondents have previously been chairs of program committees
of the ICSE Technical Research track,
another 5 have been ICSE General Chairs, and
yet another 8 have been PC chairs of other ICSE article
tracks such as SEIP, NIER, SEIS, SEET.
So at least 29 respondents (54\%) would be considered \emph{very}
senior and many of the others were similarly accomplished.

The interviews took very different routes, touching all of the
stimulus statements or only some,
often jumping back and forth between them,
refering or not refering to the respondents' own research, 
refering or not refering to things heard at the current ICSE,
and so on.
When it seemed appropriate given the flow of the interview
I would add another question near the end along the lines of:
\IV{Are ICSE and the contributions it is seeing well aligned with
    what's important for bringing software engineering forward?}.
My actual formulation often picked up statements the respondent
had made before.
26 of the interviews ended up having this section.

I analyzed the responses mostly via
Open Coding \cite[II.5]{StrCor90}.

Respondent quotes will be attributed to respondent pseudonyms 
chosen according to the names of the recording files: 
\R{394} to \R{488} (with gaps).
I sent the resulting text (that at the time was still part of
the full article about the overall study) to all respondents,
asking for feedback. 
None of the feedback I received commented on the parts presented here.

\section{Results: Is ICSE well aligned with what's important?}\label{IA}


I will first describe how respondents reported an overall positive,
negative, or mixed attitude on this question and then go into
two recurring topics in more detail.

\subsection{Overall attitudes}

Only some respondents found ICSE to be well aligned:
\begin{itemize}
\item \rQuote{464}{we're going in the right direction.}
\item \rQuote{452}{I think it's a lot better aligned than people give it
  credit for.
  [...] But as somebody who works in a company and 
  has a part-time academic role...we're 10 people here from my company, 
  all wanting to find out the latest research}.
\end{itemize}
This is the smallest group of the three.

Many respondents found ICSE not to be aligned:
\begin{itemize}
\item \rQuote{455}{No. ICSE papers are dealing with rather small problems.}
\item \rQuote{467}{Doesn't matter what's done at ICSE. Nobody else pays
  attention.}
\item \rQuote{460}{No comment.}.
\end{itemize}
The latter respondent then explained that he
had a quite negative attitude in this regard.

About half the respondents found ICSE to be partially aligned.
Some examples of those attitudes:
\begin{itemize}
\item \rQuote{483}{I am quite disturbed, almost frustrated, by the fact that we
  separate academics and practitioners.}
\item \rQuote{471}{There's always room for improvement.}
\item \rQuote{485}{In some ways I think it is but in some ways we're 
    looking at the too-immediate stuff.}
\item \rQuote{436}{my main interests are development of correct software. [...]
  [But a] lot of interest in ICSE these days is on studying how to
  increase effectiveness of software engineers themselves.} 
\end{itemize}

So by-and-large my respondents are not quite happy with the contributions
seen at ICSE today.

From the more specific comments of those critical types,
two interesting topics emerge.

\subsection{Topic 1: Some ICSE research has low relevance}

About a third of the respondents who were critical regarding
the usefulness of the ICSE contributions specifically commented
on frequently low relevance.

Some statements on this topic take an ICSE-centric perspective:
\begin{itemize}
\item \rQuote{459}{There are some topics that get too much attention without
  giving much results.}
\item \rQuote{453}{We focus too much on numbers and small details.}
\end{itemize}

Most, however, take a contributor-centric perspective:
\begin{itemize}
\item  \rQuote{430}{there is a lot of research under the lamp-post.}
\item \rQuote{472}{[Academics] are working on very academic problems.}
\item \rQuote{452}{[Much work is done because it's easy, not because it's
  important. But] that's human nature.}
\item \rQuote{481}{Some kinds of research are easier to do than others and
  nowadays there are certain kinds that are especially easy to do.}
\end{itemize}
The ``mining software repository'' type of works was mentioned several times
as a representative in this latter group.

Summing up, a sizeable subgroup of the critical respondents find the
relevance dimension of what ICSE considers high quality in a submission
ought to get more attention.

\subsection{Topic 2: Much relevant should-be ICSE research does not appear}

A similarly-sized group described research that they consider to have
\emph{high} relevance but that they feel is not often accepted at ICSE.
Three explanations are offered.

First, the technical orientation of ICSE discourages much relevant work 
on certain human or societal dimensions of SE:
\begin{itemize}
\item \rQuote{462}{You need to understand how people develop software to
  understand how you can help them. [...] 
  That's probably a bit underappreciated at the moment.}
\item \rQuote{445}{We try to compensate by having other non-technical parts or
  lesser parts either with less prestige or smaller papers or more focus
  to try and patch the [lack of societal focus] in the main part.}
\end{itemize}

Second, researchers lack the motivation to attack difficult problems
(as we saw already above):
\begin{itemize}
\item \rQuote{430}{To a large extent, for various reasons, academics are very
  opportunistic in their research.
  So the research is not driven by the importance of problems.
  In fact most academics don't know or don't define the problems well,
  because they don't have the collaboration with industry or
  the knowledge of the specific domains where software engineering 
  is happening.}
\item \rQuote{472}{I hope that we can get some reality transfer to some of the
   academics so they start working on problems that really make 
   a difference.}
\end{itemize}


Third, papers on some topics are no longer accepted because of how
evaluation is currently understood:
\begin{itemize}
\item \rQuote{485}{I tend to get very cynical about some of the things that
  get accepted, because I think it often doesn't reward the deepest things,
  it's more the things that you can quantify.
  I think it is good that we have moved in that direction [...] [, but]
  I wonder sometimes we've pushed too far.}
\item \rQuote{438}{Software architecture research was really exciting 20 years
  ago.
  As software engineering shifted with more emphasis on evaluation, 
  that research got harder to do. [...] 
  So over time software engineering research has shifted over to places where
  evaluation is easy,
  like testing or program analysis.}.
\end{itemize}

A few statements may fit into neither of these bins, e.g.
\rQuote{468}{A lot of ICSE has been fixated on building systems right. [...]
   maybe we need more research on accepting failure as a norm}.

Summing up, a sizeable subgroup of the critical respondents find that
some types of relevant research are overly rare at ICSE.

\section*{Acknowledgments}

I thank my interviewees for their time and explanations.

\bibliographystyle{IEEEtran}
\bibliography{special,agse}

\end{document}